%% file: paper.tex
\def\n{\noindent}
\input{macros.tex}
\documentclass[12pt,epsfig]{iopart}
\usepackage{epsfig}
\begin{document}
\title{Phase Stability in 3$d$-5$d$ (NiPt and CuAu) and 3$d$-4$d$ (NiPd and CuAg) Systems}
\author{\bf Durga Paudyal\footnote{email: dpaudyal@bose.res.in} and Abhijit Mookerjee\footnote{email: abhijit@bose.res.in}}
\address { S.N. Bose National Centre for Basic Sciences,
JD Block, Sector 3, Salt Lake City, Kolkata 700098, India}
\date{\today}

\begin{abstract}
We show the differences in the stability of 3$d$-5$d$ (NiPt and CuAu) 
and 3$d$-4$d$ (NiPd and CuAg) alloys arise mainly due to relativistic corrections. The 
magnetic properties of disordered NiPd and NiPt alloys also differ 
due to these corrections which lead to increase in 
the separation between $s$-$d$ bands of 5$d$ elements in these alloys. 
For the magnetic case we analyze the results in terms of splitting of  majority 
and minority spin $d$-band centers of the 3$d$ elements. We further examine 
the effect of relativistic corrections to  the pair energies and order disorder transition 
temperatures in these alloys. The magnetic moments and Curie temperatures 
have  also been studied along with the short range ordering/segregation effects 
in NiPt/NiPd alloys. 
\end{abstract}


\section{Introduction}

It is well known at the level of standard chemistry that the main chemical 
difference between pairs of 4$d$ and 5$d$ transition elements is the relativistic 
contraction of the valence $s$ and p states relative to the $d$ and $f$ States. 
Recently Wang and Zunger \cite{kn:zunger} studied ordered 3$d$-5$d$ (NiPt and CuAu) and 3$d$-4$d$ 
(NiPd and CuAg) alloys and pointed out the effect of relativistic corrections in the 
formation energies in these alloys. In this communication we shall  
provide a quantitative, electronic structure analysis of 
 these corrections  in both ordered as well as disordered phases of 
these alloys, and demonstrate its consequences on phase stability. 
We shall show, via a first-principle calculation, that in binary alloys of 
the late 3$d$-5$d$ inter-metallics, the 
 3$d$-5$d$ coupling is dominant. This effect results from the 
relativistic up-shift of the 5$d$ band, which brings it closer to the 3$d$ band 
of the other element, significantly enhancing 3$d$-5$d$ hybridization. In addition, the 
relativistic $s$ orbital contraction significantly reduces the lattice constant 
of the 5$d$ element, thus lowering the size mismatch with the 3$d$ element. This reduces 
the strain energy associated with packing 3$d$ and 5$d$ atoms of dissimilar sizes 
onto a given lattice. Both the enhanced $d$-$d$ hybridization and the reduced 
packing strain are larger in 3$d$-5$d$ inter-metallics than in 3$d$-4$d$. This explains 
why the 3$d$-5$d$ alloys CuAu and NiPt have negative formation energies and 
thus form stable ordered alloys, whereas the analogous isovalent 3$d$-4$d$ alloys 
CuAg and NiPd, made of elements from the same columns in the periodic
table, have positive formation energies and thus either phase separate or 
remain mostly in disordered phases. 
Simple arguments, such as atomic size-mismatch or electro-negativity differences, 
do not explain these two different behaviours. The constituent elements 
in the stable NiPt and CuAu alloys have larger atomic size mismatch than the unstable 
NiPd and CuAg. Likewise, the stable NiPt has a smaller electro-negativity difference then 
unstable CuAg.  

Our calculation of pair energies shows that the inclusion of relativistic 
effect in the electronic structure calculation is often important in order to get the correct 
ordering behaviour seen experimentally.
It is experimentally known that NiPt and CuAu are ordering at low temperatures, CuAg
is segregating and NiPd remains disordered but with a tendency toward short-ranged clustering.
To obtain the correct ordering tendency for NiPt and CuAu we need to carry out scalar
relativistic calculations. For CuAg  the non-relativistic calculations do show
the correct segregating behaviour. However, scalar relativistic calculations are quantitatively more
accurate. For NiPd we have to carry out calculations on the disordered alloy with
short-ranged clustering effects included, so as to give the correct magnetic moment per atom.     

From experiments the difference in the magnetic properties of disordered NiPd 
and NiPt alloys does not seem to be obvious as both Pd and Pt have the 
same number of valence electrons. Earlier works \cite{kn:par80,kn:cab70,kn:dah85} on the 
magnetic properties of NiPd and NiPt alloys used parametrized local 
environment models to describe the magnetism 
in NiPd and NiPt alloys. These models incorporated the changes 
induced due to the chemical environment as well as the magnetic environment. 
The present study is intended to improve our understanding of the reasons which   
lead to differences in the magnetic properties of disordered NiPd and NiPt 
alloys and determine  the effect of chemical, as well as  
magnetic, environments from a first principles approach. 
In an earlier communication \cite{kn:durga2} we  have  pointed out  
that environmental effect is important in NiPt alloys and single-site mean field approximations 
like the coherent potential approximation fail to predict the correct 
tendency in magnetic moments. In this paper our emphasis is on NiPd, in which we have 
found that short-ranged segregation in an otherwise disordered alloy
actually enhances ferromagnetic behaviour at par with the experimental predictions.
  
The differences in the magnetic properties of NiPd and NiPt alloys 
are also dictated by the electronic structure of 4$d$ Pd and 5$d$ Pt atoms 
and the subsequent hybridization of these states with the $d$ states of  Ni atoms. Since relativistic corrections are  
more important for heavier elements, the differences in the electronic 
structure of Pd and Pt atoms are mainly due to relativistic effects.  

\section{Theoretical and Computational methods}
For ordered structures we have performed the total energy density functional calculations.
The Kohn-Sham equations were solved in the local density approximation (LDA) with von
Barth-Hedin (vBH) \cite{kn:vbh} exchange correlations. 
The calculations have been performed in the basis of tight binding linear muffin-tin orbitals in the
atomic sphere approximation (TB-LMTO-ASA) \cite{kn:ajs}-\cite{kn:ddsam} including
combined corrections. Two sets of calculations have been performed one scalar relativistic through inclusion 
of mass-velocity and Darwin correction terms and another without. The k-space integration was carried 
out with 32$\times$32$\times$32 mesh resulting 2601 k points for tetragonal primitive structures in the
irreducible part of the corresponding Brillouin zone. The convergence of the total energies with respect
to k-points have been checked. 

As we know when the alloy is formed the elemental solids are deformed 
from their equilibrium lattice constants ($a^{0}_A$, $a^{0}_B$) to the 
lattice constants ($a$) of 
final alloy. Therefore in the alloy formation there are two types of 
formation energies. One is elastic formation energy which is given as:
\[
\Delta H_{elast} = x[E_A(a)-E_A(a^{0}_{A})]+(1-x)[E_B(a)-E_B(a^{0}_{B})]
\]
and another is chemical formation energy
\[
\Delta H_{chem} = E(A_xB_{1-x};a)-xE_A(a)-(1-x)E_B(a)
\]

\n where x is the concentration of one of the constituents.

The sum of these formation energies is the conventional alloy formation energy.
\begin{equation}
\Delta H = \Delta H_{elast} + \Delta H_{chem} 
\end{equation}

For stability arguments, we start from a completely disordered alloy. 
Each site $R$ has an occupation
variable $n_R$ associated with it. For a homogeneous perfect disorder
$\langle n_R\rangle = x$, where $x$ is the concentration of one of the
components of the alloy. In this homogeneously disordered system we now
introduce fluctuations in the occupation variable at each site : $\delta x_R =
n_R - x$. Expanding the total energy in this new configuration about the
energy of the perfectly disordered state we get :

\begin{equation}
E(x) \eq  E^{(0)}\plus \sum_{R=1}^{N} E_{R}^{(1)}\ \delta x_{R} \plus
\sum_{RR'=1}^{N} E_{RR'}^{(2)}\ \delta x_{R}\ \delta x_{R'} \plus \ldots
\label{eq:eq1}
\end{equation}

\n  The coefficients $E^{(0)}$  , $E_R^{(1)} \ \ldots $  are  the  effective
renormalized cluster interactions.
$E^{(0)}$  is the energy of the averaged disordered
medium.
The renormalized {\sl pair interactions} $ E_{RR'}^{(2)}$ express the correlation
between concentration fluctuations at two sites and are the most dominant quantities for the
analysis of phase stability.   In the series expansion, we will  retain  terms  only up  to  pair
interactions. Higher
order interactions may be included for a more accurate and
complete description. 

The total energy of a solid may be separated into two terms : a
one-electron band contribution $E_{BS}$ and the electrostatic contribution $E_{ES}$
The renormalized cluster interactions defined in (\ref{eq:eq1})
should, in principle, include both $E_{BS}$ and $E_{ES}$
contributions. Since the renormalized cluster interactions
involve the difference of cluster energies, it is usually assumed
that the electrostatic terms cancel out and only the band
structure contribution is important. Such an
assumption which is not rigorously true, has been shown to be
approximately valid in a number of alloy systems \cite{kn:heine}.
\n Considering only band structure contribution, it is easy to see that 
the effective pair interactions may be written as :
\begin{equation}
E_{RR'}^{(2)} \eq E_{RR'}^{AA}+E_{RR'}^{BB}-E_{RR'}^{AB}-E_{RR'}^{BA} 
\end{equation}
We have computed these pair energies using augmented space recursion 
with the TB-LMTO  Hamiltonian coupled with orbital peeling which allows us 
to  compute  configuration averaged  pair-potentials  directly, without  
resorting  calculations involving small differences of large total energies.
The details of this method is given in our previous paper \cite{kn:psm} and the references 
therein. 

For the calculation of order disorder transition temperature we have used Khachaturian's 
concentration wave approach in which the the stability of a solid solution
with respect to a small concentration wave of given wave vector $\vec k$ is guaranteed as long as
$ k_BT + V(\vec k)\ c(1-c) > 0 $.
Instability of the disordered state sets in when :
\begin{equation}
k_B~T^{i} + V(\vec k)\ c(1-c) = 0
\end{equation}
$T^{i}$ is the instability temperature corresponding to a given 
concentration wave disturbance.  $V(\vec k)$ is the Fourier transform of pair energies and c is 
the concentration of one of the constituent atoms. The details are give in our previous paper \cite{kn:psm} and 
references therein.

The anti-phase boundary energies between L1$_0$ and L1$_2$ structures and their
corresponding superstructures A$_2$B$_2$ and D0$_{22}$ \cite{kn:kanamori} are :
\begin{equation}
\xi = - V_2 + 4~V_3 - 4~V_4~,
\end{equation}
for $\xi \gt 0$ $L{1_2}$ and $L{1_0}$ are the stable structures  at
concentration 25 $\%$ and 50 $\%$ while for $\xi \lt 0$, the stable superstructures are $DO_{22}$ and $A_2B_2$.

Our magnetic calculations are based on the generalized ASR technique \cite{kn:durga2},\cite{kn:as}-
-\cite{kn:ppb}.
The Hamiltonian in the TB-LMTO minimal basis is sparse and therefore suitable for the
application of the recursion method introduced by Haydock \etal \cite{kn:hhk}.
The ASR allows us to calculate the configuration averaged Green functions.
It does so by augmenting the Hilbert space spanned by the TB-LMTO basis
by the configuration space of the random Hamiltonian parameters. The configuration average is expressed {\sl exactly}
as a matrix element in the augmented space.  A generalized form of this methodology is capable of
taking into account the effect of short range order. Details are given in our previous \cite{kn:durga2} paper and references therein.

For the treatment of the Madelung potential, we follow the procedure suggested by Kudrnovsk\'y \etal \cite{kn:kd}
and use an extension of the procedure proposed by Andersen \etal
\cite{kn:ajs}.
 We choose the atomic sphere
radii of the components in such a way that they preserve the total volume on the
average and the individual atomic spheres are almost charge neutral. This ensures
that total charge is conserved,  but each atomic sphere carries no excess
charge. In doing so, one needs to be careful about the sphere overlap which should be under
certain limit so as to not violate the
atomic sphere approximation.

To calculate the Curie temperature T$_{C}$ we have used the Mohn-Wolfarth (MW) procedure \cite{kn:mw} :
$$  \left(\frac{T_{C}}{T_{S}}\right)^2+\frac{T_{C}}{T_{SF}} \mns 1 \eq 0$$
where,
T$_S$ is the Stoner  temperature calculated from the relation
$$
\langle I(E_F)\rangle  \int_{-\infty}^{\infty}\ dE\ N(E)\ \left(\frac{\partial f}{\partial E}\right) 
\ =\ 1
$$
$\langle $I(E$_{F}$)$\rangle$ is the concentration averaged Stoner parameter 
, $N(E)$ is the density of states per atom per spin of the paramagnetic state\cite{kn:gun}
and $f(E)$ is the Fermi distribution function.
The spin fluctuation temperature $T_{SF}$ is given by,
\[
T_{SF} \eq \frac{m^2}{10k_{B} \langle \chi_{0}\rangle}
\]
$ \langle \chi_{0} \rangle$ is the concentration weighted exchange enhanced spin susceptibility at 
equilibrium and $m$ is the averaged magnetic moment per atom.
$\chi_{0}$ is calculated using the relation by Mohn \cite{kn:mw} and Gersdorf
\cite{kn:ger}:

\[\chi_{0}^{-1} \eq  \frac{1}{2\mu_{B}^2}\left(\frac{1}{2N^\uparrow(E_{F})}\pls
\frac{1}{2N^\downarrow(E_{F})} \mns I\right)\]

$N^\uparrow(E_{F})$ and $N^\downarrow(E_{F})$ are
the spin-up and spin-down partial density of states per atom at the Fermi level for each species in the alloy.

In these calculations one also needs to be very careful about the convergence of our procedure.
Errors can arise in the augmented space recursion because one can
carry out only finite number of recursion steps and then terminate the continued fraction using available
terminators. We ensure that the recursion is carried out for sufficient number of steps so that the errors
in Fermi energy, moments of the density of states and magnetic moment remain  within a prescribed window.

The formulation of the augmented space recursion used for the calculation in the
present paper is the energy dependent augmented space recursion in which the disordered Hamiltonian
with diagonal as well as off-diagonal disorder is recast into an energy dependent Hamiltonian having only diagonal disorder.
We have chosen a few seed points across the energy spectrum uniformly, carried out recursion on those points
and spline fit the coefficients of recursion through out the whole spectrum. This enabled us to carry out
large number of recursion steps since the configuration space grows significantly less faster for diagonal
as compared with off diagonal disorder. Convergence of physical quantities with recursion steps have been
discussed in detail earlier by Ghosh \etal \cite{kn:gdm,kn:sdgthesis}.

\section{Results and Discussion}
\subsection{Calculations on ordered alloys}
\subsubsection{Formation energies :}
\begin{table}
\caption{Formation energies in mRyd/atom. The values shown in the brackets are without
relativistic corrections.}
\begin{center}
\begin{tabular}{|c|c|c|c|}
\hline
Alloy system & $\Delta H_{form}$ & $\Delta H_{elastic}$ & $\Delta H_{chemical}$ \\
\hline
NiPd (this work)        & 3.54 (5.38)      & 17.05 (18.13)  & -13.51 (-12.75) \\
Wang and Zunger \cite{kn:zunger}        & 3.63 (6.22)      & 19.83 (21.05)  & -16.20 (-14.83) \\
\hline
NiPt (this work)        & -9.17 (4.44)     & 22.22 (31.48)  & -31.39 (-27.04) \\ 
Wang and Zunger \cite{kn:zunger}        & -6.26 (8.17)     & 29.74 (40.38)  & -36.00 (-32.21) \\
\hline
CuAg (this work)        & 6.21 (8.13)      & 16.30 (17.49)  & -10.09 (-9.36)  \\
Wang and Zunger \cite{kn:zunger}        & 7.51 (9.34)      & 18.74 (19.66)  & -11.23 (10.32)  \\
\hline
CuAu (this work)        & -5.97 (10.72)    & 19.20 (28.99)  & -25.17 (-18.27) \\ 
Wang and Zunger \cite{kn:zunger}        & -3.64 (12.16)    & 27.43 (35.13)  & -31.07 (-22.97) \\
\hline
\end{tabular}
\end{center}
\end{table}

In table 1, we show the calculated formation energies of the L1$_0$ structure of NiPd, 
NiPt, CuAg and CuAu alloys calculated relativistically (including mass velocity and 
Darwin correction but with out spin orbit couplings) as well as non-relativistically. 
The calculations were performed taking the same lattice parameters that were calculated 
by Wang and Zunger \cite{kn:zunger} and shown in their paper. 
The relativistically calculated formation energies (in mRyd/atom) are 3.54, -9.17, 6.21 
and -5.97 for NiPd, NiPt, CuAg and CuAu. We see the clear ordered alloy formation trend
of CuAu and NiPt as contrasted with the phase-separating or disordering trend of CuAg and NiPd.
To gain better insight into those trends, we have decomposed the total formation 
energies into chemical formation energy and elastic formation energy. The elastic 
energy of formation is the energy needed to deform the elemental solids A and B from 
their respective equilibrium lattice constants and to the lattice constants
of the final AB alloy. Since a deformation of equilibrium structures is 
involved, the chemical energy of formation is simply the difference between the 
(fully relaxed) total energy of the alloy and the energies of the de-
formed constituents. In general the elastic formation energy is positive and 
that of chemical formation energy is negative. The sum gives the conventional 
definition of alloy formation energy and the system is stable only if this
formation energy is negative. 
This clearly shows lower  
the volume-deformation energy of the constituents enhances the (negative) chemical 
formation energy giving rise to the the possibility of forming  a stable ordered alloy. 

Table 1, shows that the relativistic effect significantly reduces the elastic energy of formation
of 3$d$-5$d$ alloys ( e.g. from 31.48 to 22.22 mRyd/atom in NiPt and from 28.99 to 19.20 mRyd/atom 
in CuAu). This effect is much smaller in the 3$d$-4$d$ systems (~e.g. from 18.13 to 17.05 mRyd/atom 
in NiPd and 17.49 to 16.30 mRyd/atom in CuAg). The reason for this can also be appreciated by inspecting 
the non-relativistically-and relativistically-calculated equilibrium lattice constants of the fcc elements 
as already shown by Wang and Zunger \cite{kn:zunger}. 

In addition to reduction in the (positive) elastic energy of formation, table 1 also shows that relativistic corrections 
enhance the (negative) chemical energy of formation (e.g. from -27.04 to -31.39 mRyd/atom in NiPt and from 
-18.27 to -25.17 mRyd/atom in CuAu). This effect is much smaller in 3$d$-4$d$ alloys (e.g. from -12.75 to 
-13.51 mRyd/atom in NiPd and from -9.36 to -10.09 mRyd/atom in CuAg). There are two effects that explain this 
relativistic chemical stabilization. First the relativistic raising of the energy of the 5$d$ state reduces 
the 3$d$-5$d$ energy difference and thus improve the 3$d$-5$d$ bonding; second the relativistic lowering the $s$  
bands and raising of the $d$ band leads to an increased occupation of the bonding $s$ bands and a decreased 
occupation of the anti-bonding $d$ band. These effects can be appreciated by band centres shown in table 2 from 
which we can see that the 5$d$ and 3$d$ bands are closer to each other in the relativistic limit than in the 
non-relativistic limit and play important role for formation energies in these alloys. From table 3, it is 
seen that the the difference in hopping integrals between 3$d$ and 5$d$ are higher than 3$d$ and 4$d$ in 
relativistic case which is the signature of higher overlap and hence the stability in NiPt and CuAu alloys 
with relativistic corrections. The larger 3$d$-5$d$ overlap 
in relativistic NiPt than in relativistic CuAu may also explain the more negative formation energy in NiPt
than in CuAu. The $d$-$d$ interaction from different sublattices in late $d$ alloy
plays a key role. Relativity results in the raising of the energy of the 5$d$ band (bringing the 5$d$ band closer to 
the 3$d$ band) and in a large charge-transfer from the anti-bonding edge of the 5$d$ band to the bonding 6s,p bands 
thus enhancing the chemical stability of the 3$d$-5$d$ alloys.

\begin{table}
\caption{Band centres (C) mRyd/atom. The values shown in the brackets are without
relativistic corrections.}
\begin{center}
\begin{tabular}{|c|c|c|c|c|}
\hline
Alloy system & Site (A/B) & $s$ orbital       & $p$ orbital     & $d$ orbital       \\
\hline
NiPd         & Ni         & -359.0 (-343.6) & 676.3 (650.5) & -217.1 (-228.0) \\
             & Pd         & -317.6 (-240.8) & 746.2 (763.1) & -321.9 (-335.0) \\
\hline
NiPt         & Ni         & -349.4 (-347.2) & 707.0 (641.4) & -210.5 (-227.9) \\
             & Pt         & -524.0 (-283.8) & 642.1 (694.0) & -324.8 (-366.3) \\
\hline
CuAg         & Cu         & -441.9 (-423.3) & 524.9 (503.6) & -315.5 (-328.2) \\
             & Ag         & -423.5 (-362.9) & 539.4 (529.6) & -481.6 (-511.6) \\
\hline
CuAu         & Cu         & -436.1 (-426.4) & 550.6 (492.9) & -307.6 (-326.8) \\
             & Au         & -597.1 (-394.5) & 485.6 (471.9) & -456.6 (-536.2) \\
\hline
\end{tabular}
\end{center}
\end{table}

Our results for formation energies are  comparable, within the error window of our
calculational method,  to the results obtained 
by Wang and Zunger \cite{kn:zunger}. 
These authors  used full potential linearized augmented plane wave method with exchange correlation 
functional of Ceperley and Alder parametrized by Perdew and Zunger \cite{kn:zunger}. They have carried 
out k space integration with 8$\times$8$\times$8 mesh resulting 60 special k points. 
On the other hand we have in our TB-LMTO calculation used von Barth and 
Hedin exchange correlation functional and we have carried out the k space integration 
with 32$\times$32$\times$32 mesh resulting 2601 special k points to ensure the 
convergence of total energy.

\subsubsection{Separation between $s$ and $d$ band centres :}

It is seen that that the phase stability in 3$d$-5$d$ alloys are brought by relativity
through its effect on heavier atoms. We know that the most dominant effect of relativity
is to lower the $s$ potential. From table 2 it is clearly seen that the energy band centre of 
Pt in NiPt and Au in CuAu is lower in relativistic case than in non relativistic case. 
The lowering of $s$ potential causes (i) the $s$-wavefunction
to contract leading to a contraction of the lattice, and increased $s$-$d$ hybridization which
results in electron transfer from $d$ to $s$. We see that the change in $s$-$d$ separation is
more in Pt and Au in NiPt and CuAu alloys than in Pd and Ag in NiPd and CuAg alloys. The $s$-$d$
separation for Pd in NiPd alloy changes from 94.3 mRy to 4.3 mRy, whereas for Pt in NiPt alloy, it
changes from 82.5 mRy to -199.2 mRy. Similarly the $s$-$d$ separation for Ag in CuAg alloy changes from
148.8 mRy to 58.1 mRy, whereas for Au in CuAu alloy, it changes from 141.7 mRy to -140.5 mRy. Thus
the contraction of the $s$ wavefunction of Pt and the subsequent $s$-$d$ hybridization  must be
responsible for reducing the size mismatch and hence reduces the strain in NiPt and CuAu alloys
giving rise to the stable structures.

\begin{table}
\caption{Hopping integrals ($\Delta$) mRyd/atom. The values shown in the brackets are without
relativistic corrections.}
\begin{center}
\begin{tabular}{|c|c|c|c|c|}
\hline
Alloy system & Site (A/B) & $s$ orbital       & $p$ orbital       & $d$ orbital     \\
\hline
NiPd         & Ni         & 176.1 (172.6) & 155.9 (152.7)   & 9.9 (9.4)     \\
             & Pd         & 182.3  (184.8)  & 186.7 (184.3) & 22.7 (21.1) \\
\hline
NiPt         & Ni         & 177.6 (167.6) & 151.5 (142.8)   & 9.4 (8.4)     \\
             & Pt         & 163.6 (175.0) & 186.4 (183.4)   & 29.8 (25.5)   \\
\hline
CuAg         & Cu         & 153.6 (150.1) & 138.2 (133.7)   & 7.2 (6.6)     \\
             & Ag         & 157.4 (157.4)   & 163.9 (160.0) & 15.7 (14.0) \\
\hline
CuAu         & Cu         & 156.0 (145.1) & 136.8 (124.8)   & 7.0 (5.9)     \\
             & Au         & 144.0 (149.1) & 166.7 (158.9)   & 22.2 (17.3)   \\
\hline
\end{tabular}
\end{center}
\end{table}

\subsection{Calculations on disordered alloys}

\subsubsection{Effective pair energies :}

In table 4, we show the effective pair energies up to fourth nearest neighbour in 
3$d$-4$d$ NiPd, CuAg and 3$d$-5$d$ NiPt, CuAu alloys.

The first nearest neighbour pair interaction in NiPt shows ordering behaviour.
Indeed, with the relativistic correction, the antiphase boundary energy indicates
a stable ordered L1$_0$ low temperature phase. The formation energy of the
L1$_0$ phase is  negative, confirming stability. However, all these results
are true only with the inclusion of relativistic corrections.

For NiPd we have a problem. Although the positive nearest neighbour pair interaction
indicates ordering tendency, even with relativistic correction, the anti-phase
boundary energy and the formation energies indicate that the ordered structure
is not stable at low temperatures. This same remains true even if we include magnetic 
effects in the pair interaction. This alloy remains disordered in low temperatures.
However, whether this is becuase of the fact that at low temperatures the atomic
mobilities are too low for ordering to proceed fast (as in AgPd, for example), one
cannot say with certainty.

The first nearest neighbour pair interaction of  CuAg shows  segregation 
behaviour which matches  with the positive value of formation energy of the 
ordered calculations. 

In CuAu alloy the pair interaction calculations without 
relativistic correction shows segregation tendency. Inclusion of these corrections
lead to the correct conclusion of an ordering behaviour.
 Our relativistic calculation in ordered CuAu shows that the 
with L1$_0$ structure has lower total energy than the A$_2$B$_2$ superstructure. 
This confirms a stable  L1$_0$ low temperature structure of CuAu. However, the
antiphase boundary energy has the wrong sign. This could be due to the fact
that in CuAu the APB energies are long ranged and more than the fourth nearest neighbour
values need to be taken.  
   
\subsubsection{Order disorder transition temperatures :}

Using these pair interactions obtained by us, we have calculated the instability temperatures 
in NiPt and CuAu alloys with relativistic corrections. For the entropy part we have
taken a simple mean-field Bragg-Williams expression.  The calculated instability 
temperature in NiPt comes out to be 1683$^o$K 
which is higher than the experimental estimate which was discussed in great detail in our previous 
paper \cite{kn:psm}. Our calculation in CuAu alloy shows an instability temperature (246$^o$K) 
slightly higher than the experimental estimate (137$^o$K). 
The Bragg-Willimas  tends to overestimate the transition temperature, consistent with our
results.

\begin{table}
\caption{Pair energies in mRyd/atom. The values shown in the brackets are without 
relativistic corrections.}
\begin{center}
\begin{tabular}{|c|c|c|c|c|}
\hline
Alloy system & v1            & v2           & v3          & v4           \\
\hline
NiPd         & 5.16 (4.56)   & 0.02 (-0.16) & 0.05 (0.09) & 0.07 (0.09)  \\
\hline
NiPt         & 10.08 (10.11) & 0.10 (0.13)  & 0.01 (0.25) & -0.24 (0.17) \\
\hline       
CuAg         & -0.63 (-0.90) & 0.09 (0.05)  & -0.02 (0.00 & 0.12 (0.10)  \\
\hline
CuAu         & 2.29 (-0.23)  & 0.19 (0.20)  & 0.06 (0.11) & 0.21 (0.17)  \\
\hline
\end{tabular}
\end{center}
\end{table}

Our calculations (with relativistic corrections) indicate that 
 order disorder transition takes place in  NiPd at around 
(812$^o$K) is slightly higher than the non relativistic one (743$^o$K). Since there 
is no experimental evidence of order disorder transition in this system, we 
only comment that this system mainly tends to remain in a disordered phase. Looking at the 
high value (457$^{o}$K) of the experimental Curie temperature we can argue that there 
magnetism should have an effect on phase stability of NiPd system   
reducing  the value of order disorder transition temperature. Our 
calculation including magnetism indeed lowered the order disorder transition 
temperature.   

Our calculations (with relativistic corrections) on CuAg shows a
 disorder to segregation transition temperature as 184$^o$K. 
. This temperature in non relativistic case is slightly enhanced 
(201$^o$K). Both the values are lower than experimental estimate (506$^o$K).  

\subsection{Magnetic calculations in NiPd and NiPt alloys}

In figure 1, we show paramagnetic density of states for $d$ bands of Ni 
in ordered NiPd and NiPt alloys. It is seen that the $d$ band of Ni is 
narrow in NiPd in comparison to NiPt which suggests that Ni in NiPd 
has higher magnetic moment than Ni in NiPt alloys. 

\begin{figure}
\centering
\psfig{file=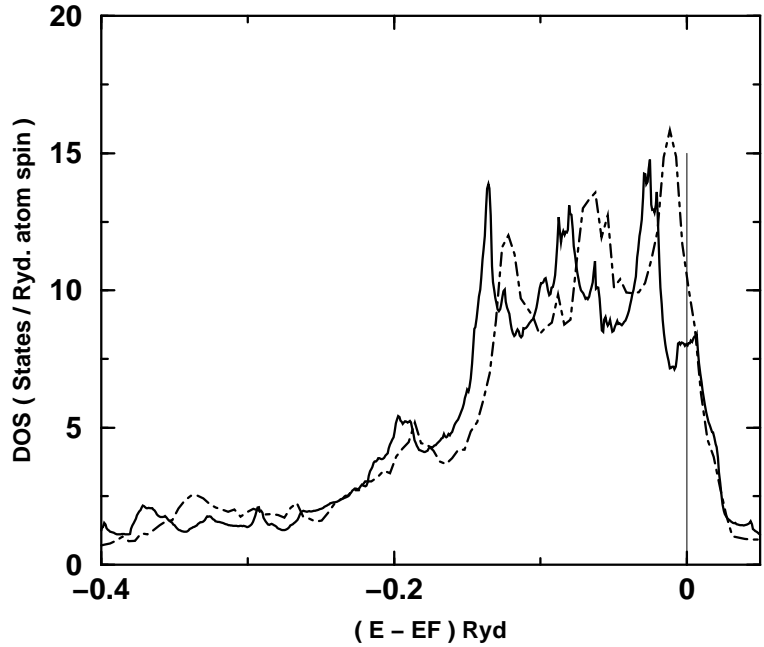,height=10cm,angle=-90}
\caption{Paramagnetic density of states for $d$ bands of Ni in ordered NiPd
and NiPt alloys.}
\label{dos_para}
\end{figure}

To understand quantitatively the differences in the magnetic properties
of these two systems, we have studied separation between majority and 
minority spin $d$ band centres, separation between $s$ and $d$ band centres 
and spin polarized density of states in these NiPd and NiPt alloys.

\subsubsection{Separation Between Majority and Minority Spin $d$-band Centers :}

The changes in the magnetic moments due 
to relativistic effects can be explained by examining the separation between majority 
spin and minority spin $d$-band centers of Ni ($\Delta C^{Ni}_{d\uparrow - 
d\downarrow}$) in NiPd and NiPt alloys. 

We note that for NiPd relativistic corrections increase the separation of $d$-bands
from 55.4 mRyd/atom to 57.3 mRyd/atom. This leads to a slight increase in the local magnetic moment
from 0.75 $\mu_B$/atom to 0.76 $\mu_B$/atom. On the other hand, in NiPt the effect is to
substantially reduce the $d$-band separation from 46.0 mRyd/atom to 21.9 mRyd/atom so that
the local magnetic moment decreases from 0.59 $\mu_B$/atom to 0.30 $\mu_B$/atom.

We  observe that the exchange-
induced splitting of the $d$-band is higher in Ni-Pd alloys for calculations 
done with and without relativistic corrections.  The higher splitting leads to an increase 
in the local magnetic moment at the Ni site. It is interesting to note that 
the inclusion of relativistic corrections produces no net change in the exchange-induced 
splitting at the Ni site in NiPd. In NiPd due to these corrections the separation 
between $d$-band centers changes from 55.4 mRy/atom to 57.3m Ryd/atom giving 
rise to 0.75$\mu_B$/atom to 0.76 $\mu_B$/atom.  On the other hand, in NiPt 
alloy we find that relativity substantially reduces the exchange-induced 
splitting at the Ni site leading to a decrease in the local magnetic moment 
of Ni. In NiPt the separation between $d$-band centers reduces from 46.0 
mRy/atom to 21.9 mRy/atom due to relativity giving rise to the corresponding 
reduction in the local magnetic moment from 0.59$\mu_B$/atom to 0.30 $\mu_B$/atom.

\subsubsection{Separation Between $s$ and $d$ Band Centers :}

It is clear that the differences in the magnetic properties of Ni-Pd and Ni-Pt alloys 
are brought about relativity through its effect on Pd and Pt atoms. We know that 
the most dominant effect of relativity is to lower the $s$ potential. The lowering of 
$s$ potential causes (i) the $s$-wavefunction to contract leading to a contraction of 
the lattice (ii) increased $s$-$d$ hybridization which 
results in electron transfer from $d$ to $s$.  We see that the change 
in $s$-$d$ separation is more in Pt than in Pd. The $s$-$d$ separation for Pd in 
NiPd alloy changes from +59.0 mRy to 7.6 mRy, whereas for Pt in NiPt alloy, it 
changes from +84.0 mRy to -199.1 mRy. Thus the contraction of the $s$ 
wavefunction of Pt and the subsequent $s$-$d$ hybridization  must be responsible for 
reducing  the local magnetic moment at the Ni site in NiPt.

\subsubsection{Spin-Polarized Densities of States :}

In figure 2 we show the spin-polarized DOS at the Ni site of 
disordered NiPd and NiPt alloys calculated with and without relativistic 
corrections. Since relativity is more important in NiPt than for NiPd, 
its effect on the DOS at the Ni site in NiPt is clearly seen. From the figure
we see the substantial differences in the density of 
electrons at Fermi level in non relativistic and scalar relativistic cases of
NiPt alloy but there is negligible difference in the case of NiPd. 

\begin{figure}
\centering
\psfig{file=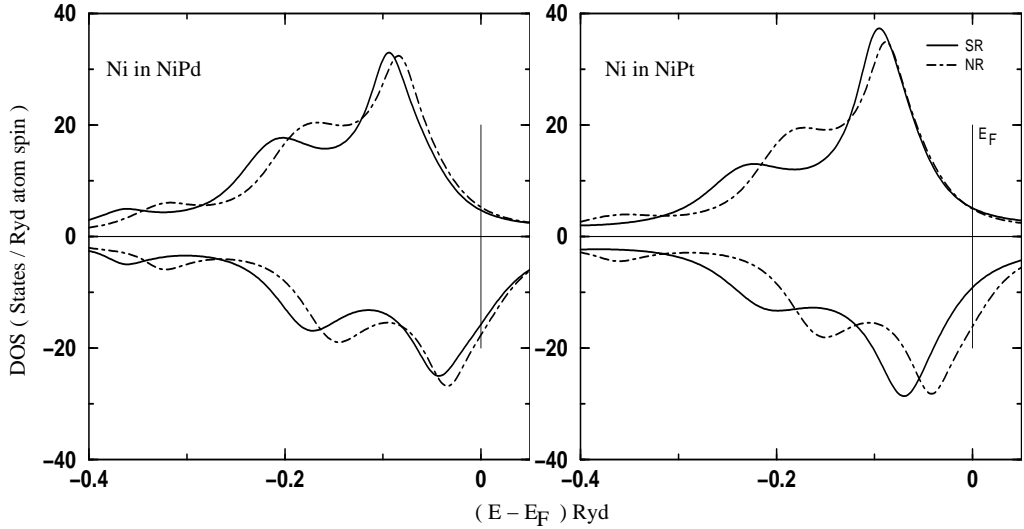,height=7cm,angle=0}
\caption{The spin-polarized densities of states of Ni, calculated
non relativistically (NR) and scalar-relativistically (SR), in
disordered NiPd and NiPt alloys.}
\label{dos}
\end{figure}

\subsubsection{Magnetic moments :}

From table 5 it is seen that magnetic moment calculated with and without 
relativistic corrections are similar in ordered as well as disordered NiPd alloys. However, 
the calculated average as well as local magnetic moments are quite different 
for ordered as well as disordered NiPt alloys with and without relativistic corrections .
We find
that the inclusion of relativity leads to a decrease in the 
magnetic moment at the Ni site by 0.29 $\mu_B$/atom in NiPt alloy system.
Our theoretically calculated disordered magnetic moment of NiPt with relativistic correction 
agrees with experimental estimates. Experimentally NiPt has got the 
effect of atomic short range order. With the inclusion of short range order effect we 
could get the magnetic moment of Ni further closer to the experimental value. The 
short range order effect in the magnetism of NiPt system is more important in the 
higher concentration of Pt (55$\%$ and 57$\%$) which we described in our previous paper 
\cite{kn:durga2}. The calculated magnetic moments of Ni in 
NiPd alloy is very low (0.76 mRyd/atom) in comparison to the diffuse scattering 
experiment (1.02  mRyd/atom). This disagreement motivated us to suspect 
the effect of short range order  on the magnetism of NiPd alloy. In order to 
check the possible short range order  effect, we have checked the variation of total energy
as a function of short range order parameter and found that
the total energy decreases as short range order parameter goes from 
negative (ordering side) to positive (segregation side) confirming this system as an segregating system.
We then checked the variation of magnetic moments as a function of SRO parameter and find
that the magnetic moments of Ni increases by appreciable fraction. The moment of Pd decreases. 
These give rise to the increase of average magnetic moment. Calculated magnetic moments 
(0.90, 0.27 and 0.59 $\mu_B$/atom for Ni, Pd and average in NiPd) including the effect of 
segregation in this system agrees closely with the corresponding experimental values 
(1.02, 0.17 
and 0.59 $\mu_B$/atom for Ni, Pd and average). 

\begin{table}
\caption{Calculated local and average magnetic moments in $\mu_B$/atom of NiPd and NiPt alloys. 
SRO and SRS denote short range order and short range segregation.}
\begin{center}
\begin{tabular}{|c|c|c|c|c|}
\hline
\multicolumn{5}{|c|}{\bf NiPd alloy}\\
\hline
Method   & System       & Ni site & Pd site & Average\\
\hline
TB-LMTO  & Ordered (SR) & 0.70    & 0.29    & 0.50\\
         & Ordered (NR) & 0.70    & 0.26    & 0.48\\ 
\hline
ASR      & Disordered (SR) & 0.76   & 0.31    & 0.54\\
         & Disordered (NR) & 0.75   & 0.28    & 0.51\\
         & Disordered (SR) with SRS & 0.90    & 0.27 & 0.59\\
\hline
Expt. \cite{kn:cab70}   & Disordered               & 1.02    & 0.17 & 0.60\\
\hline
\multicolumn{5}{|c|}{\bf NiPt alloy}\\
\hline
Method   & System       & Ni site & Pt site & Average\\
\hline
TB-LMTO  & Ordered (SR) & 0.31    & 0.16    & 0.23\\
         & Ordered (NR) & 0.65    & 0.22    & 0.44\\
\hline
ASR      & Disordered (SR) & 0.30   & 0.11    & 0.20\\
         & Disordered (NR) & 0.59   & 0.21    & 0.40\\
         & Disordered (SR) with SRO & 0.27 & 0.14 & 0.21 \\ 
\hline
Expt. \cite{kn:par80}& Disordered   & 0.28 & 0.17 & 0.22 \\
\hline
\end{tabular}
\end{center}
\end{table}

\subsubsection{Curie Temperature :}
We have applied the Mohn-Wolfarth model to calculate the curie temperature as explained 
in the theoretical and computational details. Our Curie temperature calculation  
for NiPt with the relativistic correction (76$^o$ K) shows closer agreement with the 
experimental value (100$^o$ K). In contrast the Curie temperature without relativistic 
correction in the electronic structure calculation comes out (199$^o$ K) to be higher than 
the experimental estimate. This again justifies the fact that the relativistic effect 
plays a significant role in the correct estimation of magnetic transitions as well as magnetic moments. 
The calculated Curie temperatures with (245$^o$ K) and without relativistic correction (199$^o$ K) 
do not differ much as in the case of magnetic moments as explained above. The slightly higher value 
in relativistic case may be due to slightly higher (57.3 mRyd/ atom) value of difference in $d$ band centres 
than non relativistic case (55.4 mRyd/ atom). These values of Curie temperatures with and without 
relativistic corrections do not actually 
match with experimental value (457$^o$ K). As we have explained above in the connection of magnetic 
moments, there is possibility of enhancement of magnetism due to the segregation tendency of this 
system. Our calculation taking into account this segregation effect through short range order 
parameter indeed shows the Curie temperature 
(345$^o$ K) closer to the experimental estimate. Therefore one can argue that in NiPd alloy system  
the atomic segregation tendency brings the strong enhancement in the magnetism.  

\section{Conclusions}
Our calculation for formation energies shows that NiPt and CuAu 
systems are stabilized by inclusion of relativistic effects. These  effects 
ensures larger $s$-$d$ hybridization by lowering $s$ orbitals and raising $d$ orbitals and 
lowers the strain and size mismatch in these alloys.  Similar calculations 
show NiPd and CuAg to be unstable and there is very little effect of relativity in
these systems. The pair interaction calculations in these systems shows NiPt and 
CuAu have L1$_0$ as the stable ground state structure as predicted from experiments. 
The positive value of first nearest neighbour pair energy in NiPd system 
even with the inclusion of relativistic effect indicates that this alloy 
tends to order, but experimentally it seems to remain disordered till low temperatures.
 Pair interaction calculation shows CuAg 
to be a  segregating system as predicted from experiments.
\vskip 0.2 cm
Relativistic corrections  ensure that the local magnetic moment of Ni is higher in NiPd than 
in NiPt consistent with experiment.  The low value of local magnetic moment on 
Ni site in NiPt is facilitated by relativistic corrections  again through lowering of the $s$ potential 
of Pt, which leads to a  contraction of the $s$ wavefunction and an increase in 
$s$-$d$ hybridization. We have obtained Curie temperature in NiPt  
reasonably comparable to the experimental estimate. Our Curie temperature calculation including short range 
segregation effect shows the enhancement in Curie temperature in the NiPd system  
, in better agreement with the experimental prediction than without it.

\end{document}

%% file: macros.tex
\def\P{$P$}

\def\eq{\enskip =\enskip}
\def\pls{\enskip +\enskip}
\def\mns{\enskip -\enskip}

  \def\ket{\vert \vert  \{ \emptyset \} \rangle}
  \def\ket2{\vert \vert \otimes \{ R \} \rangle}

\def\.#1{\mathaccent 95#1}
\def\^#1{\mathaccent 94 #1}
\def\~#1{\mathaccent "7E #1}

\def\plus{\enskip +\enskip}

\def\eq{\enskip =\enskip}
\def\pls{\enskip +\enskip}
\def\mns{\enskip -\enskip}

  \def\ket{\vert \vert  \{ \emptyset \} \rangle}
  \def\ket2{\vert \vert \otimes \{ R \} \rangle}

\def\gt{\; > \;}
\def\lt{\: < \:}
\def\be{\begin{equation}}
\def\ee{\end{equation}}